\documentclass[12pt]{article}

\begin{document}

 \title{Constant field strengths on $T^{2n}$}
 \author{Jan Troost \thanks{ troost@tena4.vub.ac.be; Aspirant FWO}
 \\ \hspace{3cm} \\ 
    Theoretische Natuurkunde \\ Vrije Universiteit Brussel \\
    Pleinlaan 2 \\ 1050 Brussel \\ Belgium }

 \maketitle
 \begin{abstract}
We  analyse field strength configurations in $U(N)$ Yang-Mills theory on $T^{2n}$
that are diagonal and
constant, extending early
work of Van Baal on $T^4$. The spectrum of fluctuations is
determined and the eigenfunctions are given explicitly in terms of
theta functions on tori. We show the relevance of the
 analysis to higher dimensional D-branes and discuss 
applications of the results in string theory.
\end{abstract}
 \newpage
\section{Introduction}
Constant field strengths were studied as solutions to $U(N)$
Yang-Mills on $T^4$ fifteen years ago in an attempt to get a
handle on the mechanism of
confinement \cite{Ho}. The spectrum of
fluctuations around the background field configuration was
explicitly determined by Van Baal in terms of theta functions on
the four torus \cite{VB}. These results were later used in the context
of string theory in, amongst others (\cite{GR}), \cite{HT} to compare the
spectrum of open strings ending on D-branes to the predictions
from Yang-Mills theory and the non-abelian Dirac-Born-Infeld
action. The number of massless torons
also played a crucial role in the black hole entropy counting in
\cite{CP}.

We extend the analysis of Van Baal to $U(N)$ Yang-Mills solutions
with constant and diagonal field strength on general
even dimensional tori. We will show how this analysis applies to
the low-energy effective action for D-branes in a flat background
and with small field strengths. Specifically, the
results are relevant for higher dimensional D-branes wrapped on
even dimensional tori, e.g. D6-branes wrapped around a six torus.
In the Yang-Mills theory,
we derive the conditions on the field strengths to have a stable
configuration. We examine the criterion to have massless bosonic modes
and  give explicitly the full set of fluctuations in terms
of theta functions on higher tori. In the dimensionally reduced
SYM theory relevant to D-branes the spectrum
of fermionic fluctuations  on $T^{2n}$
can also  be determined  and index theory is used
to check the number of fermionic massless modes.

The article starts out with dimensionally reducing   $D=9+1$ ${\cal N}=1$
$U(N)$ Yang Mills to the low-energy 'non-relativistic' D-brane
action. We delimit the gauge field backgrounds we will study.
Next, we compactify the theory on an even dimensional torus and
concentrate on the dependence of the fluctuations on the internal
coordinates. Here we extend the analysis of Van Baal
straightforwardly to higher tori.
Moreover we derive the spectrum of scalar and 
fermionic fluctuations in the given background and
discuss the zeromodes of the configurations in detail.
 Finally, we situate some
of the applications in string theory in the context of our
systematic analysis and indicate new applications.

 \setcounter{equation}{0}
 \section{Reduced Action}
Van Baal studied a constant diagonal field strength configuration
for $U(N)$ Yang-Mills on a four torus \cite{VB}. We will later look
at $U(N)$ Yang-Mills on a general even dimensional torus, but we will
first show how this fits in the Yang-Mills
approximation to D-brane actions. We will closely follow the
analysis and notation of \cite{VB} and
\cite{CP} in the next few sections. The
bosonic part of the $D=9+1$ ${\cal N}=1$ $U(N)$ Yang Mills action can be
written in terms of the ten dimensional field strength $G_{\mu \nu}$ 
  $ (\mu = 0, \dots , 9) $,
\begin{eqnarray}
 G_{\mu \nu} &=& \partial_{\mu} B_{\nu} - \partial_{\nu} B_{\mu}
+ i [ B_{\mu}, B_{\nu}] \nonumber
\end{eqnarray}
as
\begin{eqnarray}
S_{9+1} &=& -\frac{1}{4} \int d^{10} x \, Tr \, G_{\mu \nu}^2. 
\nonumber
\end{eqnarray}
Reduced to $p+1$ dimensions the action becomes $( \alpha = 0,
\dots , p $ and $ m = p+1, \dots,  9 ) $:
\begin{eqnarray}
S_{p+1} &=& { -\frac{1}{4} \int d^{p+1} x \, Tr \,  \left(
G_{\alpha \beta}^2 - [ \phi_m, \phi_n ]^2
                          + 2 (\partial_{\alpha} \phi_m + i
                          [B_{\alpha}, \phi_{m} ] )^2 \right) }
\end{eqnarray}
where all fields depend only on the coordinates $x^{\beta}$ and
the scalars and gauge fields  are :
\begin{eqnarray}
\phi_{m} &=& B_m (x^{\beta}) \nonumber \\ B_{\alpha} &=& B_{\alpha}
(x^{\beta}) .
\end{eqnarray}
The dimensionally reduced gauge transformations read:
\begin{eqnarray}
B & \rightarrow & \Omega B_{\alpha} \Omega^{-1} - i \Omega
\partial_{\alpha} \Omega^{-1} \nonumber \\ \phi & \rightarrow & \Omega
\phi_m \Omega^{-1}, \label{transformation}
\end{eqnarray}
with $\Omega \in U(N)$.

\section{Background and fluctuations}
In this section we determine the action for the fluctuations
around a general diagonal and constant background field strength
$G^0$. We fix notations as follows:
 \begin{eqnarray}
 B_{\alpha} &=& B^0_{\alpha} + A_{\alpha} \nonumber \\
 G_{\alpha \beta}^0 &=& \partial_{\alpha} B_{\beta}^0 - \partial_{\beta} B_{\alpha}^0
 + i [ B_{\alpha}^0, B_{\beta}^0]  \nonumber \\
 D_{\alpha} &=& \partial_{\alpha} + i [ B_{\alpha}^0, . ]  \nonumber       \\
 G_{\alpha \beta} &=& G_{\alpha \beta}^0 + F_{\alpha \beta} \nonumber \\ 
 F_{\alpha \beta} &=& D_{\alpha} A_{\beta} - D_{\beta} A_{\alpha}
                       + i [ A_{\alpha}, A_{\beta} ] 
 \end{eqnarray}
where  the gauge field
fluctuations are denoted $A$, with corresponding field strength 
$F$, and we have defined a background covariant derivative $D$.
 We choose      the background gauge fixing condition
$ D_{\alpha} A^{\alpha} =0  $. When the background gauge field and
field strength are diagonal and constant (solving the classical
equations of motion), we find the following action for the
fluctuations:
 \begin{eqnarray}
 S&=& - \frac{1}{4} \int d^{p+1} x Tr (  {G^{0 }_{\alpha,\beta}}^2 +
      - 2 A_{\alpha} D^2 A_{\alpha} - 4 i A_{\alpha} [ G^0_{\alpha \beta},
      A_{\beta} ] \nonumber \\
  & &    - 2 \phi_{m} D^2 \phi_m
      + 2 i (D_{\alpha} A_{\beta} - D_{\beta} A_{\alpha}) [ A_{\alpha},
      A_{\beta}] + 4 i \phi_m D_{\alpha} [ \phi_m, A_{\alpha}]   \nonumber \\
      & & - [A_{\alpha}, A_{\beta}]^2 - 2 [ A_{\alpha}, \phi_m]^2 - [\phi_m, \phi_n ]^2
      \label{action}
 \end{eqnarray}
 It will be convenient to
expand the fluctuations in a Lie algebra basis for $U(n)$, namely,
$(T_i)_{ab}= \delta_{ia} \delta_{ib}$ and   $(e_{ij})_{ab}=
\delta_{ia} \delta_{jb}$ for $ i \ne j$:
 \begin{eqnarray}
 A_{\alpha} &=& a^{i}_{\alpha} T_i + b^{ij}_{\alpha} e_{ij}  \nonumber \\
 \phi_m &=& c_m^i T_i + d^{ij}_m e_{ij} \label{spl}
 \end{eqnarray}
 From the reality properties of the gauge fields and the scalars we find
that $a$ and $c$ are real, and that $b^{ij}_{\alpha} = b^{ji
\ast}_{\alpha}$ and $ d^{ij}_{m} = d^{ji \ast}_{m}$ are complex.

 \section{Compactification}
 Now we will focus on our main interest. Consider
$2n$ spatial dimensions of the D-brane to be wrapped on a torus of
dimension $2n$ with radii $R_{\hat{\alpha}} =
\frac{L_{\hat{\alpha}}}{2 \pi}$, where $(\hat{\alpha}= 1, \dots, 2n )$.
In what follows, we only consider
non-trivial field strengths in these directions. Magnetic flux
quantization\footnote{We restrict to configurations where the branes are
wrapped only once around, for example, the odd cycles.
 Generalizing this to wrapping more than once these cycles 
 changes the quantization condition. 
 See for instance \cite{HT} \cite{CP}.}
and the fact that the background field strength is
diagonal implies then that we can write our background in terms of
the integers $n^i_{\hat{\alpha} \hat{\beta}}$:
 \begin{eqnarray}
 G_{\hat{\alpha} \hat{\beta}}^0 &=&
 2 \pi \frac{n^{i}_{\hat{\alpha} \hat{ \beta}}}{L_{\hat{\alpha}} L_{\hat{\beta}}}
 T_i.
 \end{eqnarray}
 We can choose the background gauge field to be:
 \begin{eqnarray}
 B_{\hat{\alpha} }^0 &=& - \pi \frac{n^{i}_{\hat{\alpha} \hat{ \beta}}
 x^{\hat{\beta} }}{L_{\hat{\alpha}} L_{\hat{\beta}}}
 T_i.
 \end{eqnarray}
Next, we substitute this form of the background into the action (\ref{action})
and concentrate
on the terms quadratic in the fluctuations. Since we will analyse
the spectrum of small fluctuations around the background, we
neglect the interactions between the fluctuations from now on. The
quadratic action is:
 \begin{eqnarray}
 S^{(2)} &=& -\frac{1}{2} \int d^{p+1} x ( a^i_{\alpha} M_0
 a^i_{\alpha} + c^i_m M_0 c^i_m  \nonumber \\
 & & + 2 \sum_{i < j} (
              b^{ij \ast}_{\alpha} (M_{ij} \delta_{\alpha \beta} - 4 \pi i
              J^{ij}_{\alpha \beta}) b^{ij}_{\beta}
              + d^{ij \ast}_m M_{ij} d^{ij}_m ) ) \label{qa}
 \end{eqnarray}
where we have defined $J$, a measure for the difference in field
strength in sector $i$ and $j$ (on brane $i$ and brane $j$), and
the mass operators $M_0$ and $M_{ij}$:
 \begin{eqnarray}
 J^{ij}_{\hat{\alpha} \hat{\beta}} &=& \frac{n^{i}_{\hat{\alpha} \hat{\beta}}
 -n^j_{\hat{\alpha} \hat{\beta}} }{L_{\hat{\alpha}} L_{\hat{\beta}}} \nonumber \\
 M_0 &=& - \partial_{\alpha}^2  \nonumber \\
 M_{ij} &=& (\frac{\partial_{0}}{i})^2 - (\frac{\partial_{2n+1}}{i})^2 - \dots
  - (\frac{\partial_{p}}{i})^2 -
 ( \frac{\partial_{\hat{\alpha}}}{i} - \pi J_{\hat{\alpha} \hat{\beta}}^{ij}
 x^{\hat{\beta}})^2. \label{Mij}
 \end{eqnarray}
 \section{The mass operators on $T^{2n}$}
{From} now on we will concentrate on the dependence of the
fluctuations on the coordinates of the torus, the internal
coordinates $x^{\hat{\alpha}}$. Specifically, we consider the
problem of diagonalizing the mass operators as operators on
$T^{2n}$.
As usual, this gives information on the spectrum of
modes living in the non-compact space. 
The quadratic action (\ref{qa}) can be analysed in each sector $ij$
separately. In the following we will leave out the indices $i$ and $j$
to simplify notation. We note that the analysis of the bosonic
fluctuations in the following sections
is valid for general $n$.

It will turn out that the mass operator (\ref{Mij}) can be 
written in a simple form.
Transforming coordinates with an element of
$O(2n)$, we can bring the difference in field strengths in sector $ij$
in the standard form :
 \begin{eqnarray}
 J_{\hat{\alpha} \hat{\beta}} &=& \left( \begin{array}{ccccc}
                0 & f_1 & 0 & 0 & 0 \\
                 - f_1 & 0 & 0 & 0  & 0 \\
                0 & 0 & \dots & 0 & 0 \\
                0 & 0 & 0 & 0 & f_n \\
                0 & 0 & 0 & - f_n & 0
                           \end{array} \right) \label{jf}
 \end{eqnarray}
 where $ f_1 \ge f_2 \ge \dots \ge f_n \ge 0 $.

  Next, we introduce
 a complex structure on the torus as follows:
 \begin{eqnarray}
 z &=& (z_1, z_2, \dots , z_n) = \frac{1}{\sqrt{2}} ( x_1 - i x_2, \dots,
 x_{2n-1} - i x_{2n} )\nonumber \\
 A &=& (A_{z_1}, A_{z_2} , \dots, A_{z_n}) \nonumber \\
 &=& \frac{1}{\sqrt{2}} (A_1+i
 A_2, \dots, A_{2n-1} + i A_{2n}) \nonumber\\
 (\partial_{z_1}, \dots , \partial_{z_n}) &=& \frac{1}{\sqrt{2}} (
 \partial_{x_1}+ i \partial_{x_2}, \dots, \partial_{x_{2n-1}} + i
 \partial_{x_{2n}} )
 \end{eqnarray}
Further, we define the  positive hermitian form $H$,
 \begin{eqnarray}
 H (z,w) &=& 2 \,  ( z_1 f_1 \bar{w_1} + \dots + z_n f_n
 \bar{w_n} ) \label{H} \\
    &=& w^{\dagger} h z \nonumber \\
 h &=& 2 \, \mbox{diag} ( f_1, \dots, f_n),
 \end{eqnarray}
 and the creation and annihilation operators
 \begin{eqnarray}
 a_k &=& \frac{1}{i } (\frac{\partial}{\partial \bar{z_k}} +  \pi
 f_k z_k) \nonumber \\
 a^{ \dagger}_k  &=& \frac{1}{i}(\frac{\partial}{\partial z_k} -  \pi
 f_k \bar{z_k}),
 \end{eqnarray}
 where $(k=1, \dots, n)$.
 Thus we can write a nice expression for the relevant non-trivial
mass operator $M_{ij}$ (\ref{Mij}):
 \begin{eqnarray}
 M &=& \{ a_k, a_k^{\dagger} \} \label{M}
 \end{eqnarray}
  Before we can diagonalize the mass operator, it is crucial to
 discuss the boundary conditions the eigenfunctions have to
 satisfy. They encode the topological data of the
 background gauge field.
 \section{Boundary conditions}
We recall that the background field strength and gauge fields were
given by:
 \begin{eqnarray}
  G^0_{\hat{\alpha} \hat{\beta}} &=& \frac{2 \pi}{L_{\hat{\alpha}} L_{\hat{\beta}}} n^i_{\hat{\alpha}
  \hat{\beta}} T_i \nonumber\\
  B^0_{\hat{\alpha} } &=& - \frac{ \pi}{L_{\hat{\alpha}} L_{\hat{\beta}}} n^i_{\hat{\alpha}
  \hat{\beta}} x^{\hat{\beta}} T_i. \label{bg}
  \end{eqnarray}
The transition functions $\Omega$ of the gauge
bundle  over the torus have to satisfy
\begin{eqnarray}
 B^{0}_{\hat{\alpha}}(x^{\hat{\beta}} + L_{\hat{\beta}}) &=& \Omega_{\hat{\beta}}
 B^{0}_{\alpha}
 (x^{\hat{\beta}}) \Omega^{-1}_{\hat{\beta}} - i \Omega_{\hat{\beta}} \partial_{\alpha}
 \Omega_{\hat{\beta}}^{-1} \nonumber \\
    &=& B^{0}_{\hat{\alpha}} (x^{\hat{\beta}}) -  \frac{ \pi}{L_{\hat{\alpha}} } n^i_{\hat{\alpha}
  \hat{\beta}}  T_i.
 \end{eqnarray}
We  choose them to be
\begin{eqnarray}
  \Omega_{\hat{\alpha}} &=& \mbox{exp} \left( -  \pi i n^i_{\hat{\alpha} \hat{\beta}}
  x^{\hat{\beta}} T_i / L_{\hat{\beta}} \right).
 \end{eqnarray} 
 The boundary conditions following from the background gauge field induced
 transition functions read, using (\ref{transformation}) and (\ref{spl}):
 \begin{eqnarray}
 a^{i}_{\alpha} ( x^{\hat{\beta}} + L_{\hat{\beta}}) &=& a^i_{\alpha} (x^{\hat{\beta}}) 
 \nonumber \\
 b^{ij}_{\alpha} (x^{\hat{\beta}} + L_{\hat{\beta}}) &=& \mbox{exp} \left(- \pi i
 n^{ij}_{\hat{\beta} \hat{\gamma}} x^{\hat{\gamma}} / L_{\hat{\gamma}} \right)
 b^{ij}_{\alpha}(x^{\beta})
 \nonumber \\
 c^{i}_{m} ( x^{\hat{\beta}} + L_{\hat{\beta}}) &=& c^i_{m} (x^{\hat{\beta}}) 
 \nonumber \\
 d^{ij}_{m} (x^{\hat{\beta}} + L_{\hat{\beta}}) &=& \mbox{exp} \left(- \pi i
 n^{ij}_{\hat{\beta} \hat{\gamma}} x^{\hat{\gamma}} / L_{\hat{\gamma}} \right)
 d^{ij}_{m}(x^{\hat{\beta}}) . \label{bc}
 \end{eqnarray}

 \section{Spectrum and eigen functions}
After the preliminary work of writing the non-trivial mass
operator $M$ in a harmonic oscillator form in terms of complex
coordinates (\ref{M}), and discussing the boundary conditions the
fluctuations have to satisfy (\ref{bc}), we determine the spectrum
and the eigenfunctions. We only discuss in detail the non-trivial
case of off-diagonal modes. Moreover, the difference between gauge
field off-diagonal modes and scalar off-diagonal modes is a mere
constant in the eigenvalues, so we can treat them in one go. 
We follow the analysis of \cite{VB} and \cite{CP}. (See \cite{H} for
an early mathematical treatment.) The
ground state $\chi_0$ of the mass operator in the off-diagonal
sector we take to satisfy the usual conditions
 \begin{eqnarray}
 0 & =& a_k \, \chi_0 \nonumber \\
 &=& \frac{1}{i} (\frac{\partial}{\partial_{\bar{z_k}}} + \frac{\pi}{2}
 h_{kl} z_l )  \chi_0,
 \end{eqnarray}
 and it has to obey the boundary conditions for the off-diagonal sector
 \begin{eqnarray}
 \chi_0 ( x^{\hat{\beta}} + L_{\hat{\beta}}) &=& e^{-\pi i n_{\hat{\beta} \gamma}^{ij}
 x^{\gamma} / L_{\gamma} } \chi_0 ( x^{\hat{\beta}} ).
\end{eqnarray}
The differential equation is immediately solved in terms of the hermitian
form $H$ (\ref{H}) and a
general holomorphic function $f$,
\begin{eqnarray}
\chi_0 (z) &=& e^{-\frac{\pi}{2} H(z,z)} f(z), \label{f}
\end{eqnarray}
but the treatment of the boundary conditions is more involved. The
boundary condition for the fluctuation $\chi_0$ implies a non-trivial
boundary
condition for the holomorphic function $f(z)$. It will be
convenient to introduce some extra machinery to write these
boundary conditions in terms of objects well-known in the
mathematical literature \cite{I} on theta functions \cite{VB}. In
terms of the hermitian form $H(z,w)$, we
define an antisymmetric form $E(z,w)$:
\begin{eqnarray}
 H(z(x),w(y)) &=& x^{\hat{\alpha}} {(-J^2)}^{\frac{1}{2}}_{\hat{\alpha} \hat{\beta}} y^{\hat{\beta}}
 + i x^{\hat{\alpha}} J_{\hat{\alpha} \hat{\beta}} y^{\hat{\beta}}  \nonumber \\
 E(z,w) &=& \mbox{Im} \, H(z,w) = \frac{1}{2i}(H(z,w)-H(w,z)) \nonumber \\
 &=& 
 x^{\hat{\alpha}} J_{\hat{\alpha} \hat{\beta}} y^{\hat{\beta}} . \label{E}
\end{eqnarray}
We introduce the notation
\begin{eqnarray}
 q&=&(q_1, \dots, q_n) \equiv \frac{1}{\sqrt{2}} (m_1 L_1 - i m_2 L_2 , \nonumber \\
 & & \dots , m_{2n-1} L_{2n-1} - i m_{2n} L_{2n} ),
 \end{eqnarray}
to write the second degree bicharacter $\alpha(q)$ in the
simple form:
 \begin{eqnarray}\alpha (q) &=&
e^{\pi i \sum_{\hat{\alpha}
< \hat{\beta}} m_{\hat{\alpha}} n_{\hat{\alpha}
 \hat{\beta}} m_{\hat{\beta}} }.
 \end{eqnarray}
These objects make it easy to write down the boundary
condition for the fluctuations $\chi$ for windings around the
torus for any number of times in different directions:
\begin{eqnarray}
 \chi_0 ( z + q) &=& \chi_0 (z) e^{-\pi i E(q,z) } \alpha (q). 
 \end{eqnarray}
 The boundary conditions the holomorphic function $f$ (\ref{f}) satisfies can
 then finally
 be written in terms of the hermitian form $H$ and the second degree
 bicharacter $\alpha$:
 \begin{eqnarray}
 f(z+q) &=& f(z) \alpha(q) e^{\pi H(z,q) + \frac{\pi}{2} H(q,q)} 
 \end{eqnarray}
Now comes the pay-off for introducing the appropriate mathematical
machinery. These holomorphic functions $f$
 are theta functions on $T^{2n}$ \cite{I}.
 They span a vectorspace of dimension $ |Pf(n_{\hat{\alpha} \hat{\beta}})| $.
This space of theta functions is the space of ground state
fluctuations around the given gauge field background. They can be
written down explicitly and we do so in the appendix.

 The higher modes are given by acting with the creation operators on the
 ground state. They automatically  satisfy the boundary conditions.
 It is clear from (\ref{M}) then that the spectrum of off-diagonal scalar field
fluctuations is given by the harmonic oscillator formula:
\begin{eqnarray}
\lambda &=& 2 \pi (\sum_{i=1}^n (2 m_i+1)f_i )
\end{eqnarray}
and after a further trivial diagonalisation (compare (\ref{qa}))
for the off-diagonal
gauge fields we get the shifted spectra:
\begin{eqnarray}
\lambda^{\pm}_k &=& 2 \pi (\sum_{i=1}^n (2 m_i+1)f_i \pm 2 f_k)
\end{eqnarray} 
 \section{Summary}
We summarize the spectrum and eigenfunctions for diagonal,
off-diagonal and gauge field and scalar fluctuations in the
following table. We use the 
notations $ e^{(z_k)} = \frac{1}{\sqrt{2}} (0,0,\dots, 1, -i,
\dots, 0,0)$ for the eigenvectors of $J$ (\ref{jf}), $V$ for
 the volume of the torus,
$i,j \in \{1, \dots , N \}; ij \in \{1, \dots, N(N-1) \}; k \in \{ 1,
 \dots, n \}; $ $p \in {\bf Z}^4; m \in {\bf N}^n;
 r_i \in \{ 0, \dots, e_i-1 \} $. All of these notations are 
 straightforward except for the components $r_i$ and $e_i$
  for which 
 we refer to the appendix. Suffice it to remark that the degeneracy of the 
 off-diagonal fluctuations is given by $Pf (n^{ij}_{\hat{\alpha}  
 \hat{\beta}})= \prod_i e_i $ in sector $ij$,
 and the space of theta functions is indexed by $ r$. We have
 moreover:
 \begin{eqnarray}
 \phi^{p,i} &=&
  \frac{1}{\sqrt{V}} e^{2 \pi i p_{\hat{\alpha}} x_{\hat{\alpha}}/ L_{\hat{\alpha}}}
 T_i
  \nonumber \\
 \phi^{{ m},{ r},ij} &=& (a_1^{\dagger})^{m_1}
 \dots (a_n^{\dagger})^{m_n} / \sqrt{m_1 ! \dots
 m_n !} \chi_{\bf r} \, e_{ij}
 \end{eqnarray}
 \begin{center}
 \begin{table}{{\it Eigenfunctions and eigenvalues}} \\ [1.5 ex]
 \begin{tabular}{|ccc|}     \hline 
         & Scalar Fluctuations &   Eigenvalues        \\ \hline 
 diagonal & $ c^i_l T_i $ &   \\ \hline            
          & $ \phi^{p,i}_l $ &
          $ \sum_{\hat{\alpha}} (\frac{2 \pi p_{\hat{\alpha}} }{L_{\hat{\alpha}}})^2 $
           \\ \hline 
 off-diagonal & $ d^{ij}_l e_{ij} $ &  \\ \hline 
  $ f_i \ne 0 $ & $  \phi^{{ m},{ r},ij}_l $ &
 $ 2 \pi (\sum_{i=1}^n (2 m_i+1)f_i ) $ \\
\hline 
& Gauge field Fluctuations &   Eigenvalues        \\  \hline   
 diagonal & $ a^i_{{\hat{\alpha}}} T_i $ &   \\    \hline 
          & $ \phi^{p,i}_{\hat{\alpha}} $ & $ \sum_{\hat{\alpha}} 
          (\frac{2 \pi p_{\hat{\alpha}} }{L_{\hat{\alpha}}})^2 $
           \\          \hline                                    
 off-diagonal & $ b^{ij}_{{\hat{\alpha}}} e_{ij} $ &  \\   \hline 
  $ f_i \ne 0 $ & $ e^{z_k} \phi^{{ m},{ r},ij} $ &
 $ 2 \pi (\sum_{i=1}^n (2 m_i+1)f_i - 2 f_k) $
        \\ 
         & $ e^{\bar{z_k}} \phi^{{m},{ r},ij} $ &
         $ 2 \pi (\sum_{i=1}^n (2 m_i+1)f_i + 2 f_k ) $  \\
\hline 
\end{tabular}\\ [0.5 ex]
 \end{table}
 \end{center} 
 We only catalogued the case where all $f^{ij}$ are different from zero
 and H is non-degenerate.
 The eigenvalues and eigenfunctions can also be classified easily in the
 other cases. 
 \section{Stability and supersymmetry}
 In this section we discuss stability, supersymmetry and the occurence of 
 massless bosonic modes in our configurations.
 From the general classification we find that there are no
  tachyonic modes when  $f_2+f_3 + \dots + f_n \ge f_1$ (for all $i,j$).
  If this condition is satisfied, stability
 of the gauge field configuration is insured, at quadratic level.
 On
 $T^4$ the condition for stability implies that the field strength
 is self-dual $(f_1=f_2)$. (Recall that we still have 
$ f_1 \ge f_2 \ge \dots \ge f_n \ge 0 $).
 The condition for stability is much more loose when
 there are more than two non-trivial field strength components. 
 Massless modes for the gauge fields occur when the equality
 $f_2+f_3 + \dots + f_n = f_1 $ is satisfied. For higher tori they appear
only in a complex combination of the gauge field
 components $A_{1,2}$. On $T^4$, stable
 configurations automatically have massless modes and they
occur for
the gauge field components $A_{1,2}$ and $A_{3,4}$.
For higher dimensions a stable configuration
 does not necessarily have bosonic massless modes.

Turning back to our starting point, we can regard the theory we
study (for $n \le 4$) 
as dimensionally reduced Super Yang-Mills theory.
The supersymmetry variation of the adjoint
fermions in our background with trivial scalars then reads:
\begin{eqnarray}
\delta \psi = F_{\hat{\alpha} \hat{ \beta}} \gamma^{\hat{\alpha}
\hat{\beta}} \epsilon.
\end{eqnarray} 
It is clear  that if $f_2+f_3 + \dots +
f_n = f_1 $, the following conditions project onto the
preserved supersymmetry parameter:
\begin{eqnarray}
-\gamma_{12} \epsilon = \gamma_{34} \epsilon = \dots =
\gamma_{2n-1 2n} \epsilon.
\end{eqnarray}
There are two other possibilities to preserve supersymmetry \footnote{We 
do not consider the case where one of the field strenghts vanishes, 
although it is easily incorporated in our framework.}
when $n=4$, namely if we have $f_1+f_4=f_2+f_3$ or if $f_1=f_2$, $f_3=f_4$.
The following projection conditions
yield the preserved supersymmetry:
\begin{eqnarray}
-\gamma_{12} \epsilon = \gamma_{34} \epsilon = 
\gamma_{56} \epsilon = - \gamma_{78} \epsilon ,
\end{eqnarray}
and 
\begin{eqnarray}
-\gamma_{12} \epsilon = \gamma_{34} \epsilon \nonumber \\ 
\gamma_{56} \epsilon = - \gamma_{78} \epsilon
\end{eqnarray}
respectively.
These last configurations do not have any bosonic massless modes.
In general,
 each projection condition halves the number of supersymmetries. 
  Note that for the four torus
   the condition for stability, namely self-duality,
coincides with the condition for preservation of supersymmetry.
For higher tori this is not the case. The space of stable gauge
field configurations is for higher tori much larger  than the space of
supersymmetric gauge field configurations.
 \section{Fermionic spectrum} 
 Up till now we ignored the fermions in our reduction of Super Yang-Mills.
 In this section we  determine the spectrum of the fermionic
 fluctuations in the theory reduced to $p+1$ dimensions, compactified 
 on $T^{2n}$, and in the background gauge field configuration (\ref{bg}).
 It will turn out that the analysis is simple once the bosonic case has 
 been treated in detail. 
 
 The  ten dimensional Majorana-Weyl
 fermions are in the adjoint representation and can
 be decomposed as
 follows:
 \begin{eqnarray}
 \psi &=& \psi^i T_i + \psi^{kl} e_{kl},
 \end{eqnarray}
where $\psi^{kl} = \psi^{lk \ast} $ and $k \ne l $.
 The relevant Dirac equation  for the fermionic modes
 is easily derived from the fermionic part of the Yang-Mills action
 quadratic in the fluctuations.
 As for the bosonic case, we can analyse the spectrum
 of the Dirac operator on  $T^{2n}$ to find the mass spectrum in the non-compact
 directions. To that end, we analyse the equation \footnote{
 The appearance of the chirality operator is to
 ensure that the non-compact and the compact part of the Dirac operator 
 commute. It does not play a crucial role
 in our analysis since it doesn't change the Dirac algebra \cite{GSW}.}: 
 \begin{eqnarray}
  ( i^{n}  \gamma_{1 \dots 2n} ) \gamma_{\hat{\alpha}} (\frac{\partial_{\hat{\alpha}} }{i} - \pi
 J_{\hat{\alpha} \hat{ \beta}}^{kl} x^{\hat{\beta}}) \psi^{kl} &=& \mu \psi^{kl} 
 \end{eqnarray} 
 We concentrate on the
 non-trivial off-diagonal components in sectors $kl$.
 The standard trick to find the spectrum of the fermionic mass operator is 
 to square it:
\begin{eqnarray}
\left( \gamma_{\hat{\alpha}} (\frac{\partial_{\hat{\alpha}} }{i} - \pi
 J_{\hat{\alpha} \hat{ \beta}}^{kl} x^{\hat{\beta}}) \right)^2 &=&
 \sum_k \{ a_k, a_k^{\dagger} \} - 2 \pi i f_1 \gamma_{12} - \dots - 2 \pi i 
f_n  \gamma_{2n-1, 2n}\nonumber
\end{eqnarray}
where we used (\ref{jf}). It is then easy to determine the spectrum of the 
fermionic mass operator by projecting onto eigenspinors of $\gamma_{2k-1,
2k}$:
\begin{eqnarray}
\lambda = \sum_k 2 \pi (2 m_k +1) f_k \pm 2 \pi f_1 \pm \dots \pm 2 \pi f_n. 
\end{eqnarray}
Implicitly, we have made use of the fact that the off-diagonal fermions 
satisfy the same boundary conditions as the bosons\footnote{We do not want
to break more supersymmetry by adding non-trivial monodromies on top of
the ones induced by the background.}. 
Drawing on the results 
in the previous sections, the eigenfunctions can then also easily be determined. 
Remark that 
the fermions all have a certain helicity associated to the
 magnetic fields in the directions $12, 34, \dots $ and each component of 
 the magnetic field is responsible for a Zeeman splitting of the energy levels. 
\section{Zeromodes and supersymmetry}
\subsection{Fermionic zeromodes}
In this section we take a closer look at the bosonic and fermionic 
zeromodes that often play a crucial role in applications. We start by
describing the fermionic zeromodes in greater detail.
It is clear from the analysis in the previous section that only 
 the following projected spinor has zeromodes:
 \begin{eqnarray}
 \psi^{kl}_{++\dots+} &=& (1+ i \gamma_1 \gamma_2) (1+ i \gamma_3 \gamma_4)
 \dots (1+ i \gamma_{2n-1} \gamma_{2n}) \frac{\psi^{kl}}{2^n} .
 \end{eqnarray}
The signs of the projection operators reverse for sector $lk$, since
the field strengths are opposite in that sector.
Explicitly, the differential equations and boundary 
conditions in sectors $kl$ and $lk$ read:
 \begin{eqnarray}
 a_i \psi^{kl}_{+\dots+} &=& 0 \label{z1} \\
 a_i^{ \ast} \psi^{lk}_{- \dots -} &=& 0 \label{z2} \\
\psi^{kl}(x^{\hat{\beta}} + L_{\hat{\beta}}) &=& e^{-\pi i n_{\hat{\beta} \gamma}^{kl}
 x^{\gamma} / L_{\gamma} } \psi^{kl} ( x^{\hat{\beta}} )
 \end{eqnarray}
 From these equations we  determine the total number of massless
fermionic modes, namely $\frac{16}{2^n} \sum_{k<l}
Pf(n^{kl}_{\hat{\alpha} \hat{\beta}})$ complex zeromodes.
In certain cases, there is a shorter route to get the number 
of fermionic zeromodes, via index theory. As a preliminary, we indicate
the $2n$ dimensional chirality of the spinors. It is clear that:
 \begin{eqnarray}
 (1- i^n \gamma_1 \dots \gamma_{2n}) \psi^{kl}_{+\dots+} &=& 0 \\
(1- (-i)^n \gamma_1 \dots \gamma_{2n}) \psi^{lk}_{- \dots -} &=& 0,
 \end{eqnarray}
such that,  for $n=2k$ even, 
the zeromodes have the same chirality in the two sectors,
while for $n=2k+1$ odd, they have opposite
 chirality. Note that this strokes with (\ref{z1}), (\ref{z2}) and the fact that 
 the complex conjugate representation of the Weyl
 representation is the original one for $ SO(4k) $, and of  different chirality
 for 
 $ SO(4k+2)$.
\subsection{Index theory check on number of zeromodes}
 Using index theory, we can learn the difference in
 number of mass\-less fermi\-onic modes of positive and negative chirality.
 If the fermionic zeromodes all have the same chirality, index theory
 predicts the total number of fermionic
zeromodes. 
 We will use the index
 theorem for the twisted spin complex on a flat manifold for the
 adjoint representation \cite{EGH} :
 \begin{eqnarray}
 \mbox{index} ( \Delta_{\pm} \times (adj),D) &=& \int_{M}  ch
 (F_{adj}) \nonumber \\ &=& \nu_{+} - \nu_{-}
 \end{eqnarray}
 where $ ch(F_{adj})= \sum_j
 Tr \frac{F^j_{adj}}{j ! (2 \pi)^j}$ is the Chern class evaluated in the adjoint
 representation and $ \nu_{\pm} $ is the number of positive, negative
 chirality zeromodes.
 In the background (\ref{bg}) the
 integral is easily evaluated on $T^{2n}$:
 \begin{eqnarray}
  \nu_{+} - \nu_{-} &=& \sum_{k l} Pf (n^{kl}_{\hat{\alpha} \hat{\beta}}).
 \end{eqnarray}
 For $d=4k+2$ the sum over all sectors is zero,
 the contributions from sector $(kl) $
 (strings going one way)
 cancelling the contribution from sector $(lk)$ (strings going the other way). 
 Indeed, from the analysis in the previous section we know that 
 the number of zeromodes of positive chirality equals the number of zeromodes
 of negative chirality in this case.
 For $d=4k$ we find, in our conventions, only
 zeromodes of positive chirality, and index theory counts
 $ \sum_{k  l}
  Pf (n^{kl}_{\hat{\alpha} \hat{\beta}})$ complex fermionic zeromodes in $d=4k$.
  Taking into account the multiplicity of the zeromodes originating
  in ten dimensions, we find
$ \frac{16}{2^n} \sum_{k < l}
  Pf (n^{kl}_{\hat{\alpha} \hat{\beta}})$ non-constant 
  complex spinor zeromode components,
  as before.
 This straightforwardly extends the
 well-known results in four dimensions  \cite{Schwarz}.
\subsection{Supersymmetry and massless modes}
The number of massless fermionic and massless bosonic modes differs 
in supersymmetric configurations, and 
at first sight it is difficult to see how they form a representation of 
supersymmetry when it is partially unbroken. Nevertheless, they do. We discuss this 
slightly puzzling feature  in this subsection.
The unbroken supersymmetry transformations rules are  given by the
dimensionally reduced formulae
of ${\cal N} =1 $ SYM in ten dimensions. In ten dimensions the formulae read: 
\begin{eqnarray}
\delta A_{\mu}^a &=& \frac{i}{2} \bar{\epsilon} \gamma_{\mu} \psi^a 
\label{us1} \\
\delta \psi^a &=& -\frac{1}{4} F_{\mu \nu}^a \gamma^{\mu \nu} \epsilon 
\label{us2}
\end{eqnarray}
Consider first the case where the unbroken supersymmetry is given by
\begin{eqnarray}
- \gamma_{12} \epsilon = \gamma_{34} \epsilon = \dots = \gamma_{2n-1 2n} 
\epsilon. \label{p1}
\end{eqnarray}
Starting out with a fermionic zeromode satisfying
\begin{eqnarray}
 \psi^{kl}_{++\dots+} &=& (1+ i \gamma_1 \gamma_2) (1+ i \gamma_3 \gamma_4)
 \dots (1+ i \gamma_{2n-1} \gamma_{2n}) \frac{\psi^{kl}}{2^n}, \label{p2}
 \end{eqnarray}
 we easily see from (\ref{us1}), (\ref{p1}) , (\ref{p2}) and $\mu =m$
 that it will never transform into a 
 massless scalar. That is consistent with the spectrum we found earlier. When
 $n \ge 3$, we find only non-trivial gauge field components in the $1$ and 
 $2$ direction. For gauge field components in the other directions,
 the projection condition on the 
 parameter $\epsilon$ and the zeromode $\psi$ make sure that the
 variation vanishes. This is
 again consistent with what we found earlier. It can easily be checked 
 that on $T^4$ a similar analysis yields  complex bosonic zeromodes in 
 directions $1,2$ and $3,4$. Moreover for the cases where the supersymmetry 
 parameter satisfies $\gamma_{12}  \epsilon = 
 -\gamma_{34} \epsilon$ and $ -\gamma_{56} \epsilon = \gamma_{78} \epsilon $ we 
 find that no bosonic zeromodes are generated, as expected.
 Finally, it can be checked using the formulae for the bosonic zeromodes 
 that the transformation rule (\ref{us2}) is always trivial (at the order
we are working). Thus, the 
 analysis of fermionic and bosonic zeromodes is perfectly consistent with 
 supersymmetry, as it should be. Of course, one could  perform 
 a similar consistency check for higher modes in the spectrum.
 \section{Applications and conclusion}
The theory we have been discussing is  the low-energy theory of 
D-branes in a trivial background and with only small ('non-relativistic') 
gauge fields excited. Indeed, string theory should be seen as the 
high-energy completion of the theories we studied. Nevertheless, the 
approximation to string theory we discussed was often used to understand 
results in string theory. We will briefly review some of the applications 
of the results that we obtained that have already been made in the 
literature and that can be coherently presented and extended
in our framework. 

In \cite{HT} the fluctuation spectrum on $T^4$ 
in the Yang-Mills approximation was 
compared to the string theory fluctuation spectrum and the role of the 
(non-abelian) Dirac-Born-Infeld action in resolving the discrepancy was 
clarified. The precise form of the non-abelian Dirac-Born-Infeld action
remained unclear. Our analysis could be useful for
further studying fluctuation spectra 
from the different points of view along the lines of \cite{HT}, for higher
branes on higher tori. Note also that the 
explicit form for a tachyonic fluctuation was used in \cite{HT} to discuss 
tachyon condensation intuitively. That discussion is now easily extended 
to higher branes.

The condition for the background gauge fields to preserve supersymmetry is 
familiar in D-brane physics, especially in its T-dual form. To repeat this 
well-known point, it is sufficient to give an archetypical example. 
 Consider the following configuration: a pair of D8-branes compactified on
 $T^8$ with a constant
 field strength on the first D8-brane $F_{12} \ge F_{34} \ge F_{56} \ge F_{78}
 \ge 0 $. When $  F_{12} = F_{34} + F_{56} + F_{78} $, 
 supersymmetry is conserved. 
 T-dualize this configuration over directions $2,4,6,8$ to obtain a
 pair of D4-branes at angles. The angles are given by the following formulae
 \footnote{We ignore some constant factors.}:
 $ F_{12} = \tan{\phi_1} $,
 $ F_{34} = \tan{\phi_2 }$,
 $ F_{56} = \tan{\phi_3 }$,
 $ F_{78} = \tan{\phi_4 }$. Working at small angle, or taking into account
 the modifications the Born-Infeld action induces in the Yang-Mills theory in
 the spirit of \cite{HT}, we find that the condition
 coincides with the well-known one for rotated branes \cite{BDL} \cite{Pol}.
 Note though that for the case $n=4$ we saw a physical 
 distinction between the case where $f_1=f_2+f_3+f_4$ and 
 $f_1+f_4=f_2+f_3$. The difference between the two cases is probably 
 related to the mechanism of the creation of a string in the D0-D8 system
 (or D4-D4 system), but we do not pursue this here.

As noted before, on higher tori, the space of stable gauge field
configurations is 
much larger than the space of supersymmetric configurations. The same 
applies therefore to all kinds of D-brane constructions in gauge theories.
The results on 
$T^4$ for the Yang-Mills theory were extensively discussed in for instance
\cite{GR} \cite{HT}. Stability of 
some special configurations on $T^6$ and $T^8$ was used in \cite{T}
to adhere zerobranes to six- and eightbranes in Matrix theory.
It is now obvious that these are 
points in a larger parameter space of stable configurations.
Note though  that we only proved stability up to quadratic order in
the fluctuations \cite{T}. It may be useful to recall
that for instance the black hole with
only D0- and D6-brane charge, which has a corresponding quadratically stable
representation in the gauge theory on the D6-brane, is in fact 
metastable  \cite{T} \cite{06}.

Another motivation for our work can be found in the following problem.
Consider  the only regular four dimensional
supersymmetric black hole that is  solely made out of
D-branes \cite{BL}. Specifically, dualize to 
the configuration where the compact part carries D6-D2-D2-D2 brane charges.
The entropy for this black hole, calculated in supergravity, was 
microscopically accounted for in \cite{BL} \cite{BLL} up to a constant factor 
which it would be interesting to determine.
We can now easily find supersymmetric configurations in the gauge theory living
on the D6-branes that represent such a black hole.  
Next, we can use the results derived in this article to calculate the dimension
of the moduli space of these configurations by counting massless states.
Although in principle this program as suggested in \cite{BL} looks sound, we 
have as yet not been able to make it work. 
Finally, we indicate that the analysis of strings stretching between branes in
Matrix theory 
is analogous (see for instance \cite{pp} )
and we believe that our systematic treatment
could be of practical use in that context too.

To sum up, we have determined the spectrum and explicit 
eigenfunctions of the fluctuations around constant and diagonal field 
strengths in $U(N)$ Yang-Mills on an even dimensional torus, extending
earlier work \cite{VB} \cite{CP} on $T^4$. We discussed 
supersymmetry and stability of these configurations, and the counting
of 
zeromodes. The analysis yields a systematic framework 
for applications in string theory.

  \vspace{1cm}

  \noindent {\bf Acknowledgments}:
It is a pleasure to thank  Ben Craps, Frederik Denef, Marc Massar,
Frederik Roose, Alex Sevrin and especially Walter Troost for 
useful
discussions, and Wati Taylor and Pierre Van Baal  for
friendly correspondence.

 \section*{Appendix: Explicit  eigenfunctions}
 To find the explicit form of the theta functions we need some more
 machinery \cite{I}. We go to a canonical (Frobenius) basis of the lattice
using  an $SL(2n,{\bf Z})$ transformation:
 \begin{eqnarray}
 n &=&
 \left( \begin{array}{cccc}
 0 & 0 & e_1  & 0 \\
        0 & 0 & \dots  & \dots \\
        0 & 0 & 0 & e_n     \\
        -e_1 &  0 & 0 & 0  \\
        \dots & \dots & 0 & 0 \\
         0 & -e_n & 0  & 0
 \end{array}             \right)
 \end{eqnarray}
 where the $e_i$ are positive integers, each $e_i$ dividing the next 
 $e_{i+1}$.
 We choose a {\bf C}-basis as follows:
 \begin{eqnarray}
 z &=& \tilde{z_1} \frac{\xi^{(n+1)}}{e_1} + \dots + \tilde{z_n}
 \frac{\xi^{(2n)}}{e_n} = U \tilde{z}
 \end{eqnarray}
 where $\xi^{\hat{\alpha}}$
 denotes the canonical basis of the lattice.
 By $X$ we denote the real vector space generated by $\xi^{n+1}, \dots,
 \xi^{2n}$.
 Since $Im H = E$
 (\ref{E}) is  zero on $X \times X$,
 we find that $\tilde{h} = U^{\dagger} h U $ is real and
 symmetric. We define then the symmetric $C$-bilinear form $S$ 
 uniquely associated with $H$
 \begin{eqnarray}
 S(z,w)= - \tilde{w}^T \tilde{h} \tilde{z} 
 \end{eqnarray}
 and the quasi-hermitian form $Q$ 
 \begin{eqnarray}
 Q(z,w)= H(z,w) + S(z,w).
 \end{eqnarray} 
 Moreover,  we need the period matrix $\tau$ defined in terms of the quasi-hermitian
 form $Q$:
 \begin{eqnarray}
 \tau_{ij} \equiv \frac{i}{2} Q(\xi^{(i)}, \xi^{(j)}). 
 \end{eqnarray}
Further definitions are required, namely the bicharacter
$e^{\pi i B(q,q)}$ with respect to
the canonical basis $\xi$:
\begin{eqnarray}
        q &=& \tilde{q_{\hat{\alpha}}} \xi^{\hat{\alpha}} \nonumber \\
        B(q,q) &=& \sum_{i=1}^{n} e_i \tilde{q_i} \tilde{q_{i+n}} 
\end{eqnarray}
and $\beta(q)$, encoding the characters $m$ and $l$ of $\alpha$:
 \begin{eqnarray}
 \alpha(q) &=& e^{\pi i B(q,q)} \beta(q)\nonumber \\ 
 \beta(q) & =& e^{2 \pi i \sum_{i=1}^{n} (m_k \tilde{q_{k+n}} - l_k \tilde{q_k)}}
 \end{eqnarray}
The thetafunctions are then explicitly given by:
 \begin{eqnarray}
 \theta(z) &=& \sum_{0 \le r_i < |e_i| } d_r \theta_r (z)\nonumber \\
 \theta_r (z) &=& \sum_{p \in Z^n} e^{\pi i (p+ e^{-1} (m+r) . \tau
 (p+e^{-1} (m+r)) + 2 \pi i (p + e^{-1} (m+r)) . (\tilde{z}+l)}
 \end{eqnarray}
 where $ e_1 | e_2 | e_3 \dots | e_n $
 and $e$ is diagonal.

  \end{document}